\documentclass{cdclass}
\usepackage{float}
\usepackage{tabularx}
\usepackage[table]{xcolor}
\newcommand{\orcidicon}{\includegraphics[width=0.32cm]{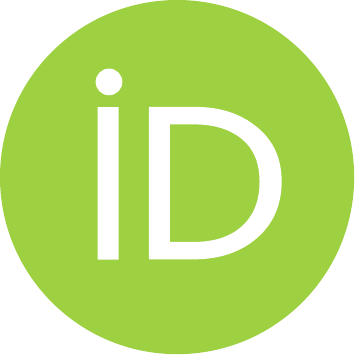}}

\begin{document}

\title{Influence of Gender Composition\\ in Pedestrian Single-File Experiments}

\author{
  Sarah Paetzke\href{https://orcid.org/0000-0003-0761-2459}{\orcidicon}\authorlabel{1} \and 
  Maik Boltes\href{https://orcid.org/0000-0001-7240-896X}{\orcidicon}\authorlabel{1} \and
  Armin Seyfried\href{https://orcid.org/0000-0001-8888-0978}{\orcidicon}\authorlabel{1, 2}
}
\authorrunning{S. Paetzke \and M. Boltes \and A. Seyfried}
\institute{
  \authorlabel{1} Institute for Advanced Simulation, IAS-7: Civil Safety Research, Forschungszentrum Jülich,
  \authortab{1} 52425 Jülich, Germany,
  \authoremail{1}{s.paetzke@fz-juelich.de}
  \and
  \authorlabel{2} Faculty of Architecture and Civil Engineering, University of Wuppertal,
  \authortab{1} 42285 Wuppertal, Germany
}

\date{2023}{21 February 2023}{}{}

\maketitle

\begin{abstract}
Various studies address the question of what factors are relevant to the course of the fundamental diagram in single-file experiments. Some indicate that there are differences due to group composition when gender is taken into account. For this reason, further single-file experiments with homogeneous and heterogeneous group compositions were conducted. A Tukey HSD test was performed to investigate whether there are differences between the mean of velocity in different density ranges. A comparison of different group compositions shows that the effect of gender can only be seen, if at all, in a small density interval. Regression analyses were also conducted to determine whether, at high densities, the distance between individuals depends on the gender of the neighboring pedestrians and to establish what human factors have an effect on the velocity. An analysis of the distances between individuals at high densities indicates that there is no effect of the gender of the neighboring pedestrians. Taking into account additional human factors in a regression analysis does not improve the model.
\end{abstract}

\keywords{Pedestrian dynamics \and single-file movement \and culture
  \and gender effect \and regression analysis}

\section{Introduction}
In recent years, there have been a number of studies that have shown that fundamental diagrams of various geometrical settings such as stairs \cite{ye_pedestrian_2021,burghardt_performance_2013}, single-file experiments \cite{cao_pedestrian_2016, chattaraj_comparison_2009, was_effects_2014}, corridors \cite{cao_investigation_2018, feliciani_empirical_2016, hu_social_2021, ren_fundamental_2019, jin_observational_2019}. or crossings \cite{cao_fundamental_2017, holl_methoden_2016} vary \cite{predtechenskii1978planning, weidmann_transporttechnik_1993, vanumu_fundamental_2017, fruin_pedestrian_1971,  zhang_pedestrian_2012, vanumu_fundamental_2017}. However, it is not only the spatial structure that creates differences. When we go look more closely at into more detail about the specific structure, it becomes clear that there are also variations depending on the experiment setup. The type of flow such as uni-, bi-, or multidirectional streams, human factors such as age, gender, height, and culture \cite{ziemer_mikroskopische_2020,cao_pedestrian_2016, ren_contrastive_2019,subaih_experimental_2020,subaih_questioning_2022, bandini_phase_2010,zhang_universal_2014,migon_favaretto_investigating_2019,nguyen_gender-based_2019,dias_pedestrians_2022}, or external factors such as restricted visibility \cite{cao_dynamic_2019}, different height adjustments due to smoke \cite{ma_experimental_2020}, motivation or instruction \cite{ziemer_mikroskopische_2020}, rhythm or background music \cite{zeng_experimental_2019,yanagisawa_improvement_2012}, or properties of human movement such as step length and frequency \cite{cao_stepping_2018,zeng_experimental_2018, wang_linking_2018, ma_pedestrian_2018, song_experiment_2013, wang_step_2018, fujita_traffic_2019} all affect the fundamental diagram.

The question of what factors are relevant to the course of the fundamental diagram in single-file experiments is not yet clear. It is also difficult to compare experiments partly owing to the combination of different human factors and partly because the measurement methods or the experimental scenarios vary, too. For instance, for one experiment, there might only be data in the low-density range whereas another experiment might also have data for the high-density range. Furthermore, it should be noted that often the problem arises that the fundamental diagrams represent a group that is homogeneous in one factor but different in terms of other factors. This problem was discussed in more detail in \cite{paetzke_influence_2022} where single-file school experiments were studied to analyze how human factors affect the fundamental diagram of pedestrian dynamics.

With respect to the effect of gender, the results of some existing studies can be summarized as follows. Subaih et al.\,\cite{subaih_experimental_2020} have shown that for densities higher than 1.0~m$^{-1}$ groups compositions homogeneous in gender lead to higher speeds than a heterogeneous group composition with alternating order. But a comparison with data from other cultures and different ages raise the question of what other factors also need to be considered. In \cite{subaih_questioning_2022}, using the data from the experiments introduced in\cite{subaih_experimental_2020}, Subaih et al.\,have shown that the headway to the front and to the back is important, too. This result suggests that the arrangement by gender has an effect on the distances between pedestrians and must be taken into account in modeling the speed-density relation. While these findings indicate a significant contribution of gender in Paetzke et al.\,\cite{paetzke_influence_2022}, it was still concluded that gender could be neglected. This analysis is based on a multiple linear regression from experiments with heterogeneous group compositions.  

To analyze these contradictory findings, further single-file experiments are performed for the present study. Four different group compositions, female, male, gender alternating, and gender random order are considered to investigate the following three hypotheses derived from the studies to date \cite{subaih_experimental_2020,subaih_questioning_2022,paetzke_influence_2022}. 
\begin{enumerate}
\item The speed-density relation depends on the gender composition of the group of test persons.
\item At high densities, the distance between individuals depends on the gender of the neighboring pedestrians.
\item The inclusion of additional human factors that were not previously included such as the weight, the exact height, and the gender of the previous pedestrian improves the multiple linear regression model developed in \cite{paetzke_influence_2022}.
\end{enumerate}

For the first hypothesis, the question is whether there are differences within the density-velocity relation between the mean values of the homogeneous and heterogeneous pedestrian group compositions when gender is taken into account. This has been tested in seven density intervals using the Tukey HSD test. For the second hypothesis, simple linear regression analysis is used to determine whether there are differences between the group compositions at high densities. For the third hypothesis, a multiple linear regression analysis is performed with different human factors. 

The second section of this paper describes the experimental setup, the measurement methods, the data preparation, and the experiments that are compared. Section 3 deals with the results and analysis of the hypotheses. A comparison of two different experiments based on the group composition is carried out and the regression analysis, which includes simple and multiple linear regression, is performed. The conclusions are presented in the last section and further research is proposed.
\section{Materials and Methods}
\subsection{Experimental setup}
\begin{figure}[H]
\centering 
\begin{minipage}[t]{0.45\linewidth} 
\centering
\includegraphics[width=7.4cm]{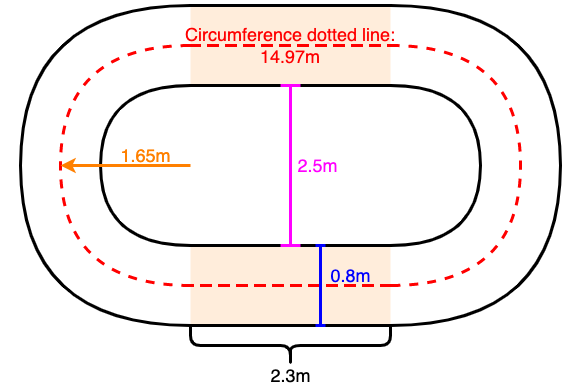}\\
a)
\end{minipage}  
\hfill 
\begin{minipage}[t]{0.45\linewidth} 
\centering
\includegraphics[width=6.cm]{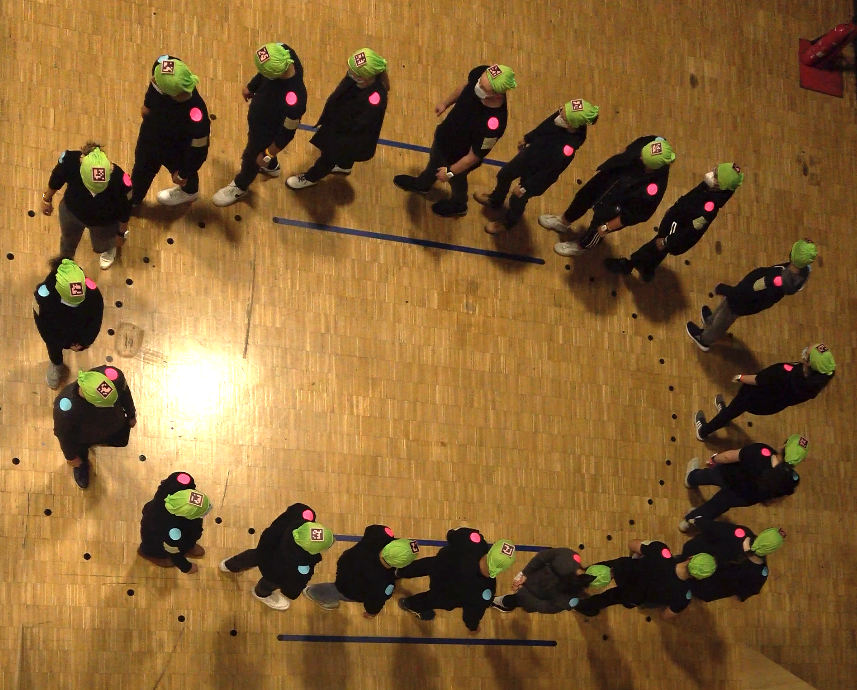}\\
b)
\end{minipage}   
\caption{Single-file experiment performed at the Mitsubishi Electric Halle in Düsseldorf, Germany in 2021. Figure a) shows the oval path with the lengths of the central line, the width of the oval path, the middle radius, the length of the straight sections, and two measurement areas highlighted in the background. Figure b) shows the experiment from above with the students wearing green caps with personal ID codes on top.}
\label{fig:geometrysetup}
\end{figure}

The subject of the present study is a one-dimensional single-file experiment performed within an experimental series \cite{DatenPaperCroMa} of the projects CroMa and CrowdDNA at the Mitsubishi Electric Halle in Düsseldorf, Germany in 2021. The oval path measurements in the experiment are total length of the central line $l=14.97$~m by width of $w=0.8$~m. The middle radius is 1.65~m while the straight sections are 2.3~m long. The two measurement areas are highlighted in the background in the sketch on the left in \autoref{fig:geometrysetup}.  

When the pedestrians' trajectories are projected into the central line of the oval, the complexity of the system is reduced to one dimension and, consequently, only the change in movement direction is taken into consideration \cite{ziemer_congestion_2016}. 
Four different group compositions were considered in the experiment. Two homogeneous group compositions with respect to gender were chosen, male (m) and female (f). A third group is a heterogeneous group composition where the male and female participants are arranged in an alternating order m, f, m, f, etc. A fourth group is also a heterogeneous group composition, but with a random structure, as the male and female pedestrians were randomly distributed in the oval, for example, m, m, f, m, f, f, etc. With both homogeneous group compositions, ten experimental runs at different global densities were performed. The global density is adjusted by $N$, the number of persons situated in the oval. For the runs, we chose $N=4, N=8, N=16, N=20, N=24, N=32, N=36$, and $N=40$. In the two experiments with heterogeneous group compositions, all test subjects participate at least once in a run at each density. This is not the case in the two homogeneous runs. The densities of the experiments with the heterogeneous group compositions correspond to the densities in the experiments with the homogeneous group compositions, but there are now a total of 25 different experimental runs instead of ten. Global densities $\rho_{gl}=N/l$ for these cases were $\rho_{gl}\in[0.27,2.67]$~m$^{-1}$.

For all parts, the runs have a duration of between two and three minutes. Two minutes were chosen for runs with $N < 32$ because, up to this density, the pedestrians have moved a long distance in the oval in the time considered.
The test persons were instructed to walk behind each other without haste or overtaking. In total, $80$ different pedestrians participated in the experiment with an equal ratio of male and female pedestrians. They all wear green caps with personal ID codes on top. These codes are used to extract the trajectories of different participants in several experimental scenarios and to assign personal information to a participant such as a gender, age, shoulder width, weight, and height \cite{boltes_automatic_2010, BOLTES2013127}. 

\autoref{table:1} shows a detailed overview of the mean values and the standard errors for age, height, weight, and shoulder width for the four different groups: female, gender alternating, gender random order, and male. The average age of the participants is between 26 and 28~years, their average heights range between 1.70~m and 1.83~m, and their average weights are between  
    76.26~kg and 92.24~kg and, lastly, their average shoulder widths are between 0.43~m and 0.49~m.
    \begin{table}[H]
      \caption{The columns show a detailed overview of the mean values and the standard errors for age, height, weight, and shoulder width for the different groups: female, gender alternating, gender random order, and male.\label{table:1}}
\scriptsize
\setlength{\tabcolsep}{5pt}
\renewcommand{\arraystretch}{1.7} 
\begin{tabular}{cccccccccc}
\toprule
     &  \textbf{Female} & \textbf{Gender alternating} & \textbf{Gender random order} & \textbf{Male} \\
    \midrule
  $\overline{age}\pm \sigma$ in $years$     & $27.12\pm 8.11$ &$27.74\pm 6.03$ & $25.97\pm 5.14$ & $26.42\pm 4.92$ \\
  $\overline{height}\pm \sigma$ in $m$                                   & $1.70\pm 0.08$ &$1.75\pm 0.09$ & $1.77\pm 0.11$ & $1.83\pm 0.07$ \\  
    $\overline{weight}\pm \sigma$ in $kg$                               & $76.26\pm 21.16$ &$92.24\pm 20.61$ & $80.95\pm 20.88$ & $88.74\pm 26.08$ \\
   $\overline{shoulder \ width}\pm \sigma$ in $m$                            & $0.43\pm 0.03$ &$0.45\pm 0.04$ & $0.46\pm 0.05$ & $0.49\pm 0.03$ \\
\bottomrule
  \end{tabular}
  \centering
\end{table}

\subsection{Measurement methods}
The individual velocity, the Voronoi tessellation, and density are calculated on the basis of the one-dimensional trajectories obtained by tracking the head from the video recording. 
For this case, $x_i(t)$ describes the position of individual $i$ at time $t$. Pedestrian $i+1$ is walking directly in front and a person $i-1$ directly behind person $i$.
The Voronoi distance $d_{V_i}(t)$ of pedestrian $i$ at time $t$ is calculated by 
\begin{linenomath}
\begin{equation}\label{eq:headway}
d_{V_i}(t)=\frac{1}{2}\cdot (x_{i+1}(t)-x_{i-1}(t)) \ ,
\end{equation} 
\end{linenomath}

which is the half of the distance between the centers of the heads $x_{i+1}(t)$ and $x_{i-1}(t)$. The density is calculated by $\rho_{i}(t)=\frac{1}{d_{V_i}}$. The individual velocity is calculated by
\begin{linenomath}
\begin{equation}\label{eq:indvelo}
v_i(t)=\frac{x_i(t+\frac{\Delta t}{2})-x_i(t-\frac{\Delta t}{2})}{\Delta t} \ .
\end{equation}
\end{linenomath}

As explained in \cite{paetzke_influence_2022}, the value $\Delta t=0.8$~s is a good assumption. The intended direction and negative velocities of the pedestrians are also included. Both straight sections of the oval are used as measurement areas.

\subsection{Data processing}
For the various experimental runs, only the data in a steady state are considered. The range was determined by the CUSUM algorithm \cite{liao_detection_2016}. To ensure the independence between two successive measurement values, such as for the velocity, autocorrelation was used to determine one value for the time gaps to be considered in an experimental run. On average, the time gap between these measurement values is about 1.38 seconds. For each group composition, approximately 3,000 data points are considered for the analysis. 

\subsection{Experiments in comparison}
The single-file experiments in Düsseldorf, Germany - performed for the present study are compared with the single-file experiments conducted by Subaih et al.\,\cite{subaih_experimental_2020, dataSubaih} at the Arab American University in Palestine. Therefore, the data from Subaih's study, already including the velocity and density, are used. In this section, only selected features of the experiment conducted by Subaih et al.\,are described. For further details, we refer to \cite{subaih_experimental_2020}. In Subaih's experiment, the measurements of the oval path are total length of the central line of $l=17.30$~m by width of $w=0.6$~m indicated by markings on the floor. The straight sections are $3.15$~m long. In total, $47$ different pedestrians participated in the experiment with 26 female and 21 male students. Their heights are within the range of 1.52~m to 1.84~m. On average, the height of men is 1.75~m and women 1.61~m. Their age is between 18 and 23~years. For the homogeneous group compositions including both male (UM) and female (UF), the number of persons situated in the oval is $N=14$ and $N=20$. For the heterogeneous group composition with a gender alternating order (UX), there are also $N=24$ and $N=30$. The global densities are $0.81$~m$^{-1}$, $1.16$~m$^{-1}$, $1.38$~m$^{-1}$ and $1.73$~m$^{-1}$, respectively. Compared to the experiments in Düsseldorf, Germany, the participants in Palestine are younger and shorter than the pedestrians in Germany. In both experiments, the experimental scenarios of the homogeneous group compositions in terms of gender were performed first. 
Furthermore, the same measurement methods are used in both experiments. 
\section{Results and Analysis}
\label{Analysis}
\subsection{Comparison of group compositions for the experiments performed in Germany}
\begin{figure}[H]
\centering
\includegraphics[width=12cm]{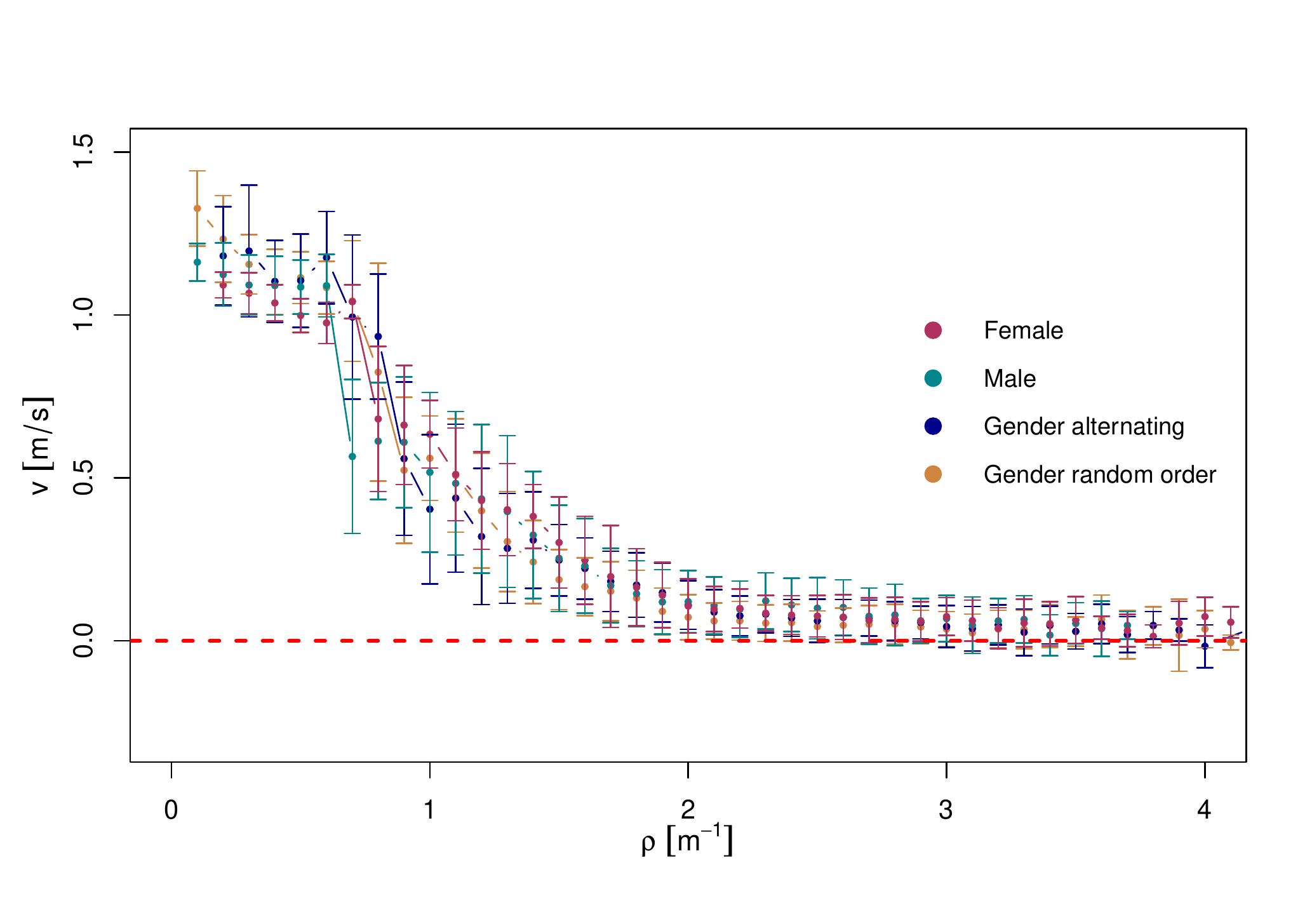}
\caption{Fundamental diagram of density vs.\,velocity for the groups female, male, gender alternating and gender random order in Germany with binned data within 0.1 intervals.}
\label{fig:FDDensityVeloBinning}
\end{figure}

\begin{figure}
\centering
\includegraphics[width=11cm]{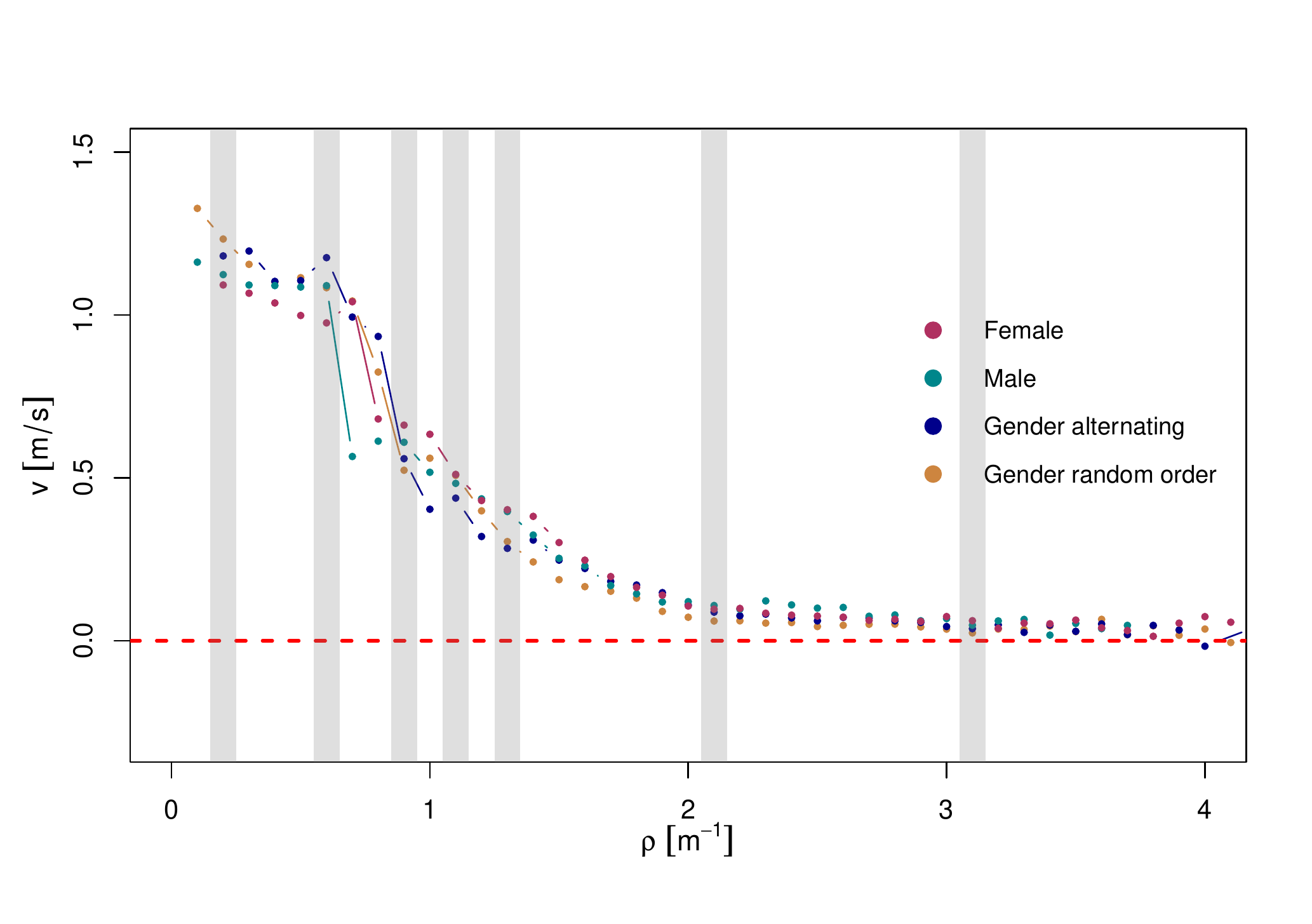}
\caption{Fundamental diagram of density vs.\,velocity fundamental diagram for the groups female, male, gender alternating, and gender random order in Germany with binned data that represent only the mean values of the velocity. Furthermore, the individual intervals $[0.15, 0.25]$, $[0.55, 0.65]$, $[0.85, 0,95]$, $[1.05, 1.15]$, $[1.25, 1.35]$, $[2.05, 2.15]$, and $[3.05, 3.15]$ are highlighted in grey areas.}
\label{fig:FDDensityVeloBinningIntervalls}
\end{figure}

In order to check the first hypothesis, we conducted an analysis for different density intervals to see whether there are systematic differences in the velocity between homogeneous and heterogeneous group compositions with respect to gender. First, a visual comparison was performed. \autoref{fig:FDDensityVeloBinning} shows a density vs.\,velocity fundamental diagram for the groups female, male, gender alternating, and gender random order in Germany with binned data so that the trend and possible differences can be seen more clearly.
The data suggest that the question of whether the fundamental diagrams correspond or not, depends on which density interval is considered. Up to a density of about 1.15~m$^{-1}$, there seems to be no systematic variation in equality and inequality between the group compositions. For densities larger than 1.15~m$^{-1}$ and smaller than 2.0~m$^{-1}$, it could be assumed that the velocity is higher for homogeneous group compositions. For densities higher than 2~m$^{-1}$, the course of the individual curves looks very similar. For a more detailed analysis, the mean values of the velocity are compared for each group composition in seven small density intervals $[0.15, 0.25]$, $[0.55, 0.65]$, $[0.85, 0,95]$, $[1.05, 1.15]$, $[1.25, 1.35]$, $[2.05, 2.15]$, and $[3.05, 3.15]$ which are highlighted in grey areas in the fundamental diagram in \autoref{fig:FDDensityVeloBinningIntervalls}. 
First, the Kolmogorov-Smirnov test was conducted to determine the velocity distribution for all group compositions and this resulted in a difference between almost all distributions in all intervals. Therefore, a statistical test, a Tukey HSD (honest significant difference) test was then performed to check whether the means are significantly different from each other. Here, all group compositions are directly compared pairwise. The test takes into account that the sample size is approximately the same. If the Tukey test shows that the p-value is larger than 0.05, this means that there is equality between the means of the observed group compositions considered. 

\begin{table}
\caption{Mean values and the standard deviation for the velocity $(\overline{v}\pm \sigma)$ in seven different density intervals for different group compositions in Germany and Palestine. Equal colors in an interval indicate equality between the corresponding groups.\label{table:2}}
\newcolumntype{C}{>{\centering\arraybackslash}X}
\begin{tabularx}{\textwidth}{CCCCCCCC}
\toprule
 &  \footnotesize{$[0.15, 0.25]$}     & \footnotesize{$[0.55, 0.65]$} &\footnotesize{$[0.85, 0.95]$} & \footnotesize{$[1.05, 1.15]$} & \footnotesize{$[1.25, 1.35]$} & \footnotesize{$[2.05, 2.15]$}  & \footnotesize{$[3.05, 3.15]$}\\
 \hline
  \scriptsize{Female, Germany} & \scriptsize{$1.07\pm 0.04$}     & \scriptsize{$1.10\pm 0.05$} & \cellcolor[HTML]{f2bdb2}\scriptsize{$0.62\pm 0.14$} & \cellcolor[HTML]{f2bdb2}\scriptsize{$0.58\pm 0.12$} & \cellcolor[HTML]{f2bdb2}\scriptsize{$0.41\pm 0.14$} & \cellcolor[HTML]{f2bdb2}\scriptsize{$0.10\pm 0.07$} & \cellcolor[HTML]{95655c}\scriptsize{$0.06\pm 0.06$} \\
   \hline
  \scriptsize{Male, Germany} & \cellcolor[HTML]{f2bdb2}\scriptsize{$1.17\pm 0.09$}     & \cellcolor[HTML]{f2bdb2}\scriptsize{$1.11\pm 0.10$} & \cellcolor[HTML]{f2bdb2} \scriptsize{$0.61\pm 0.19$}& \cellcolor[HTML]{bd8074}\scriptsize{$0.48\pm 0.23$} & \cellcolor[HTML]{f2bdb2}\scriptsize{$0.44\pm 0.21$} & \cellcolor[HTML]{f2bdb2}\scriptsize{$0.11\pm 0.09$} & \cellcolor[HTML]{95655c}\scriptsize{$0.06\pm 0.08$} \\
   \hline
  \scriptsize{Gender alternating, Germany} & \cellcolor[HTML]{f2bdb2}\scriptsize{$1.17\pm 0.16$}     & \cellcolor[HTML]{f2bdb2}\scriptsize{$1.14\pm 0.15$} &\cellcolor[HTML]{fdeeeb}\scriptsize{$0.74\pm 0.24$} & \cellcolor[HTML]{bd8074}\scriptsize{$0.43\pm 0.22$} & \scriptsize{$0.30\pm 0.19$} & \cellcolor[HTML]{f2bdb2}\scriptsize{$0.10\pm 0.07$} & \cellcolor[HTML]{95655c}\scriptsize{$0.05\pm 0.06$} \\
   \hline
 \scriptsize{Gender random order, Germany} & \scriptsize{$1.26\pm 0.14$}    & \cellcolor[HTML]{f2bdb2}\scriptsize{$1.11\pm 0.07$} &\cellcolor[HTML]{f2bdb2}\scriptsize{$0.54\pm 0.30$} & \cellcolor[HTML]{f2bdb2}\scriptsize{$0.55\pm 0.15$} & \scriptsize{$0.35\pm 0.16$} & \scriptsize{$0.07\pm 0.07$}  & \cellcolor[HTML]{95655c}\scriptsize{$0.04\pm 0.06$}\\
\hline
\hline
 \scriptsize{Female, Palestine} &     &   &\cellcolor[HTML]{fdeeeb}\scriptsize{$1.14\pm 0.01$}  &\cellcolor[HTML]{fdeeeb}\scriptsize{$0.80\pm 0.11$} & \cellcolor[HTML]{fdeeeb}\scriptsize{$0.68\pm 0.07$} &  \\
\hline
 \scriptsize{Male, Palestine} &     &   &\cellcolor[HTML]{fdeeeb}\scriptsize{$0.94\pm 0.20$}  &\cellcolor[HTML]{fdeeeb}\scriptsize{$0.81\pm 0.13$} & \cellcolor[HTML]{fdeeeb}\scriptsize{$0.74\pm 0.13$}  &  \\
 \hline
 \scriptsize{Gender alternating, Palestine} & &     &\cellcolor[HTML]{fdeeeb}\scriptsize{$1.00\pm 0.18$}  &\cellcolor[HTML]{fdeeeb}\scriptsize{$0.70\pm 0.14$} & \cellcolor[HTML]{f2bdb2}\scriptsize{$0.50\pm 0.18$} & \cellcolor[HTML]{f2bdb2}\scriptsize{$0.11\pm 0.05$} \\
\bottomrule
\end{tabularx}
\end{table}
\unskip
\autoref{table:2} shows the mean values and the standard deviation for the velocity of each group in Germany and Palestine in all seven intervals. The results of the experiments performed in Palestine by Subhai et.al.\,will be discussed in the next subsection in the comparison.

For every interval, group comparisons where the p-value is larger than 0.05-and so their mean values are therefore equal-are shown in the same color. 
In the first four intervals in the range from 0.15~m$^{-1}$ to 1.15~m$^{-1}$, both the group with the highest speed changes and the group compositions that are rated as equal by the Tukey test change from interval to interval. First, the two means are significantly equal for female and gender random order. In the second interval, the test results show equality between male, gender alternating, and gender random order, then between female, male, and gender random order and, finally, in the fourth interval, between female and gender random order as well as male and gender alternating. Consequently, there is no systematic equality or inequality between different group compositions in the low-density regime. In the density range between 1.25~m$^{-1}$ and 1.45~m$^{-1}$, the mean values of the homogeneous group compositions are equal to or higher than those of the heterogeneous group compositions. Between 1.45~m$^{-1}$ and approximately 2.15~m$^{-1}$, the two means are significantly equal for all group compositions except gender random order. For the group gender random order, the velocity is the lowest of all groups. At a density of 2.15~m$^{-1}$ and above, there is equality for all group compositions. 
\begin{figure}[H]
\centering 
\begin{minipage}[t]{0.49\linewidth} 
\centering
\includegraphics[width=6.5cm]{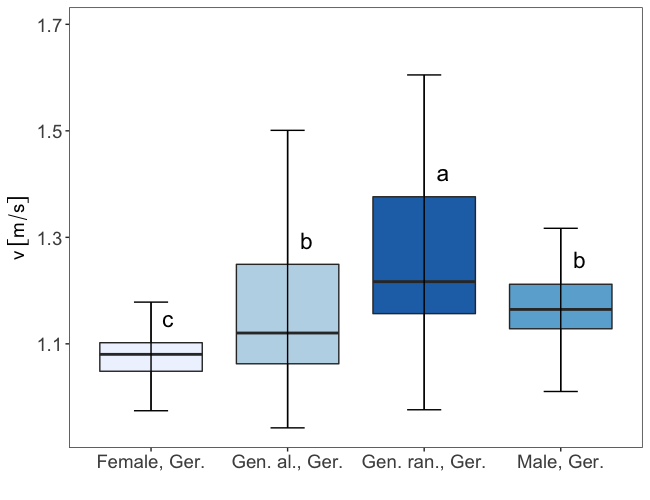}\\
a)
\end{minipage}  
\hfill 
\begin{minipage}[t]{0.49\linewidth} 
\centering
\includegraphics[width=6.5cm]{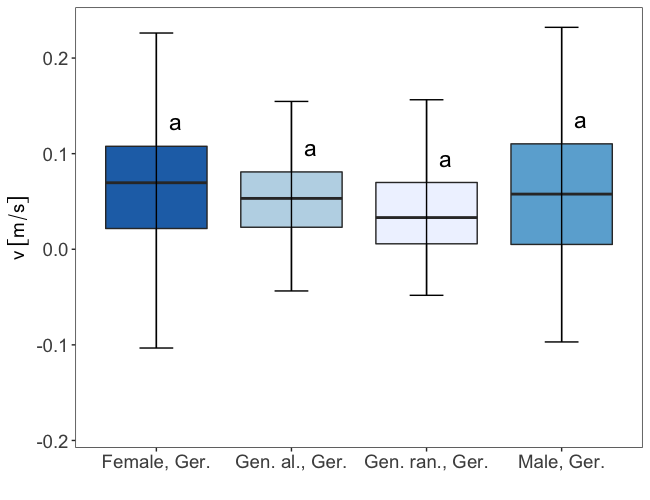}\\
b)
\end{minipage}  
\caption{Boxplots for the velocity for different group compositions in Germany. Equal letters above the boxes indicate equality of the mean values of the velocity between the corresponding groups. The first interval $[0.15, 0.25]$ is represented by a) and the last one $[3.05, 3.15]$ by b).}
\label{fig:BoxplotGermany1und7}
\end{figure}
In addition to \autoref{table:2}, the boxplots in \autoref{fig:BoxplotGermany1und7} also illustrate the results of the Tukey test. The left-hand figure a) shows the first interval $[0.15, 0.25]$ at which there is only equality between male and gender alternating groups and the right-hand one b) shows the last interval $[3.05, 3.15]$ at which all group compositions are equal. This equality or inequality is indicated by the letters above the boxes. The same letters represent equality between group compositions. In addition, the boxplots show the minimum and maximum values, the median within the box, and the lower and upper quantiles, the boundaries of the box, and of the individual group compositions. The color of the box is darker, the higher the median is.

\subsection{Comparison with the experiments performed in Palestine}
\begin{figure}[H]
\centering 
\begin{minipage}[t]{0.49\linewidth} 
\centering
\includegraphics[width=7.2cm]{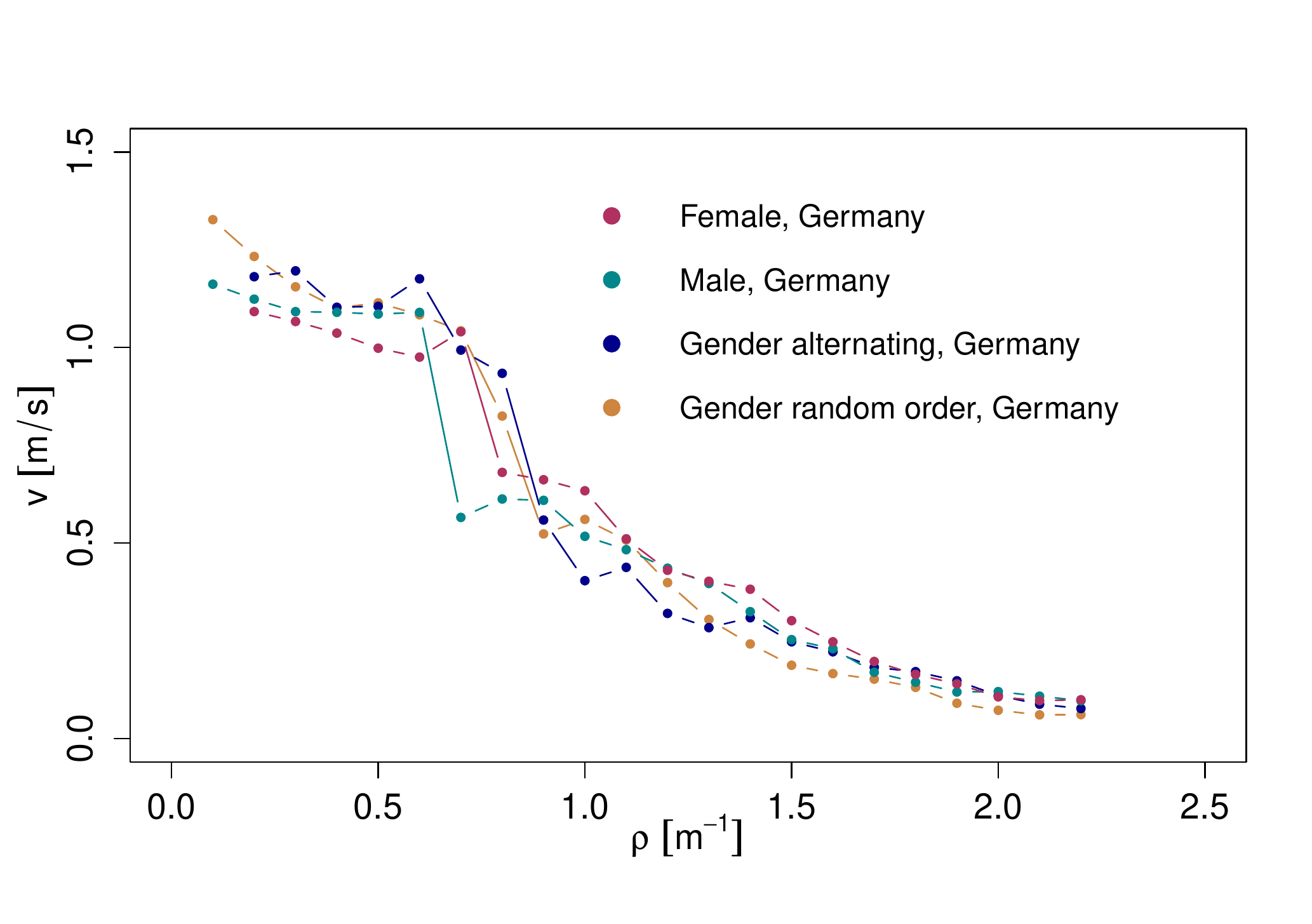}
a)
\end{minipage}  
\hfill 
\begin{minipage}[t]{0.49\linewidth} 
\centering
\includegraphics[width=7.2cm]{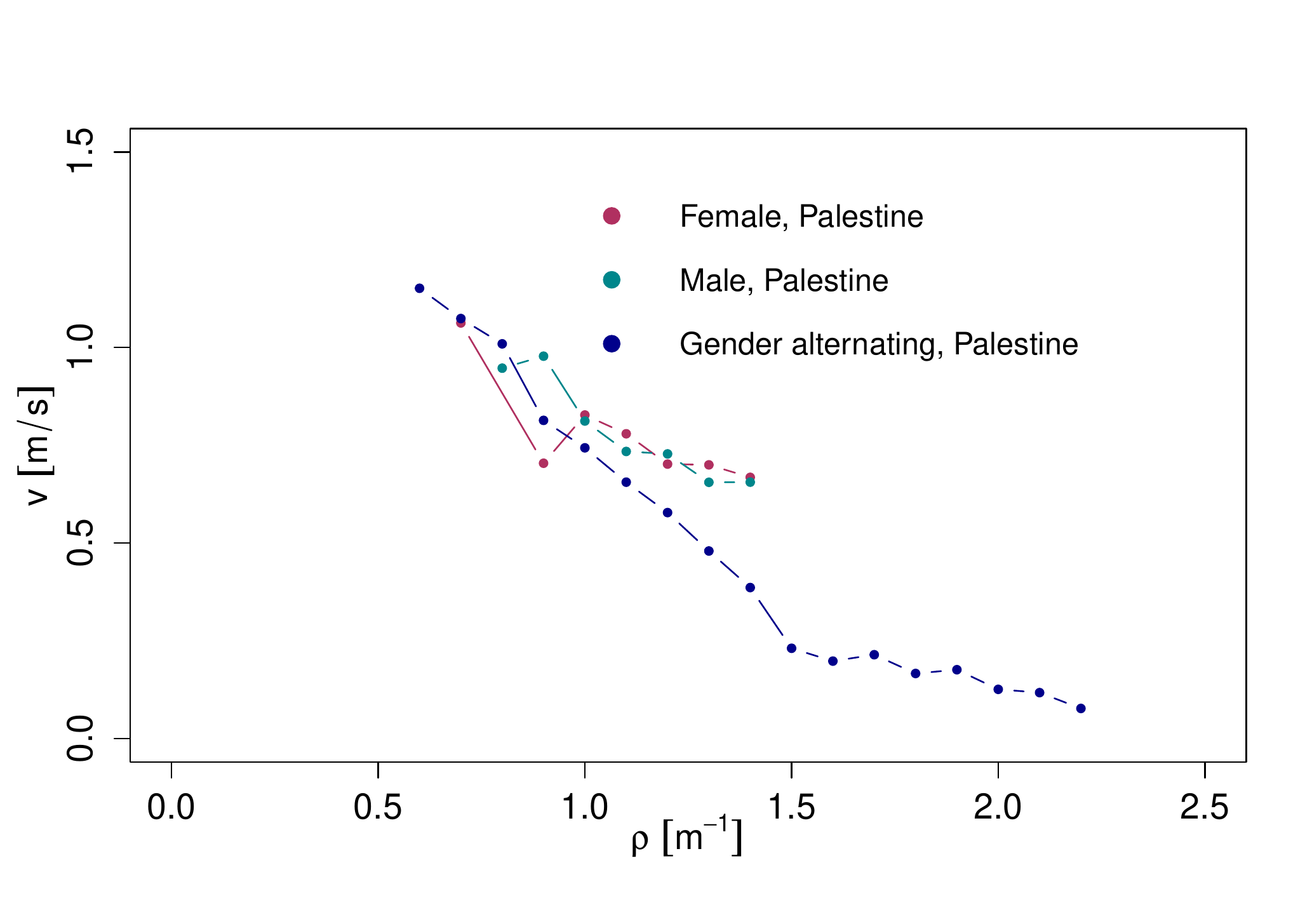}
b)
\end{minipage}  
\caption{Relation of density and velocity for the experiments in a) Germany and in b) Palestine by binned data of the mean values of the velocity for different group compositions.}
\label{fig:FDDensityVeloBinningRudina}
\end{figure}

In this section, data from the experiments in Palestine \cite{subaih_experimental_2020, dataSubaih} are compared with those in Düsseldorf, Germany \cite{DataSingleFileCroMa}. First, the binned data are plotted in a fundamental diagram up to a density of 2.5~m$^{-1}$ (see \autoref{fig:FDDensityVeloBinningRudina}). The left-hand diagram illustrates the values for Germany and the right-hand one those for Palestine. 
A visual comparison based on \autoref{fig:FDDensityVeloBinningRudina} shows that in the density interval of [0.6, 1.5]~m$^{-1}$, the mean velocity is higher in the experiments performed in Palestine than in Germany. For densities higher than 1.5~m$^{-1}$, the heterogeneous group in Palestine approaches the mean speed of the groups in Germany. When we only consider the data from Palestine, we see that the heterogeneous group composition in Palestine is faster for densities less than 0.8~m$^{-1}$. With increasing density, the homogeneous group compositions of males and females show higher means than the heterogeneous group composition. 

\begin{figure}[H]
\centering
\includegraphics[width=13cm]{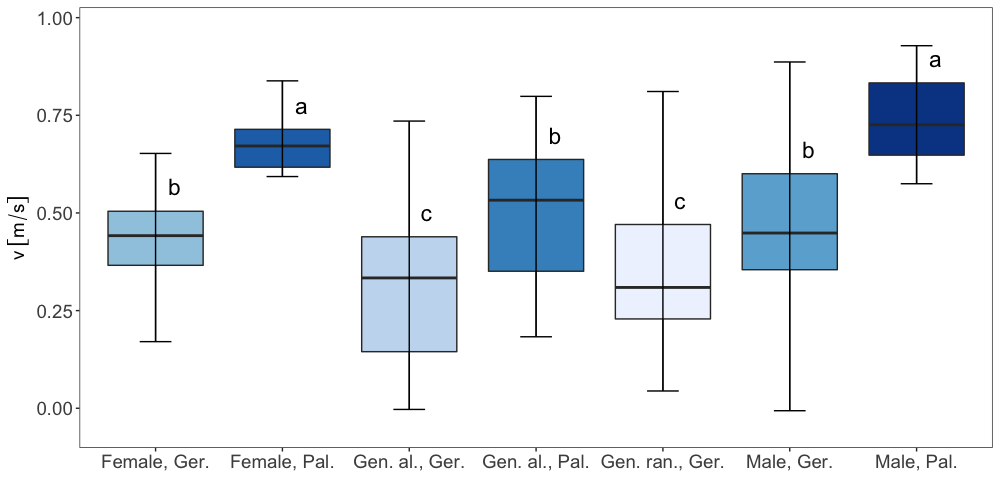}
\caption{Boxplots for the velocity for different group compositions in Germany and Palestine in the interval of $[1.25, 1.35]$~m$^{-1}$. The same letters above the boxes indicate equality for the mean values of the velocity between the corresponding group compositions.}
\label{fig:BoxplotPalestine}
\end{figure}

To enable us to compare the data of the experiments in Germany and Palestine based on a statistical test, the Tukey HSD test was used again. The corresponding mean values and the standard deviation of the velocity are shown in \autoref{table:2}. For the experiments performed in Palestine, only data for the intervals $[0.85, 0.95]$~m$^{-1}$ to $[2.05, 2.15]$~m$^{-1}$ are available. 
Up to a density of approximately 1.0~m$^{-1}$, the means are significantly equal for all group compositions in Palestine, as well as for gender alternating in Germany, and for the group compositions of female, male, and gender random order in Germany. In the interval $[2.05, 2.15]$~m$^{-1}$, the two means are significantly equal between german group compositions and between all group compositions in Palestine. Above a density of 1.25~m$^{-1}$, the heterogeneous group composition in Palestine approaches the group compositions in Germany. Only in the interval $[1.25, 1.35]$~m$^{-1}$ are the homogeneous group compositions equal in Palestine and in Germany and have a higher mean value of the velocity than the heterogeneous groups in the corresponding countries. In other density intervals, there are no significant differences in Palestine and no systematic differences in Germany. \autoref{fig:BoxplotPalestine} shows the results of the Tukey test for the comparison between Germany and Palestine in the interval of $[1.25, 1.35]$~m$^{-1}$. 
\subsection{Gender differences in distances of neighboring pedestrians}
\label{Gender differences in distances of neighbouring pedestrians}
A simple linear regression analysis for the speed-distance relation of each individual $i$ was conducted to determine whether the distance between individuals depends on the gender of the neighboring pedestrians at high densities. The model is: 
\begin{linenomath}
\begin{equation}\label{eq:simple_model}
v_{i,l}=\beta _{0_i}+\beta _{1_i} \cdot (d_{V})_{i,l}+\epsilon_{i,l} \ ,
\end{equation} 
\end{linenomath}
where $l=1,...,n_i$ and $n_i$ is the number of individual observations. $v_{i,l}$ is the individual velocity, the predicted variable, $\beta_{0_i}$ is the intercept, $\beta_{1_i}$ is the regression coefficient, $(d_{V})_{i,l}$ is the Voronoi distance as an independent variable, and $\epsilon_{i,l}$ describes the random experimental error.
For a good adjustment, the values $\beta_{0_i}$ and $\beta_{1_i}$ need to be estimated. This results in the following equation: 
\begin{linenomath}
\begin{equation}\label{eq:simple_estimation}
\hat{v}_{i,l}=\hat{\beta}_{0_i}+\hat{\beta}_{1_i} \cdot (d_{V})_{i,l} \ .
\end{equation} 
\end{linenomath}
When we transform formula \ref{eq:simple_estimation} for $\hat{v}_{i,l}=0$, the minimum Voronoi distance for each individual results:
\begin{linenomath}
\begin{equation}\label{eq:dmin}
(d_{V})_{i,min}=-\frac{\hat{\beta}_{0_i}}{\hat{\beta}_{1_i}} \ .
\end{equation} 
\end{linenomath}
$\hat{\beta}_{1_i}$ can be interpreted as the reaction time for acceleration and braking. 

In \autoref{table:3}, the columns provide an overview of the mean values and the standard deviation for $(d_{V})_{i,min}$ and $\hat{\beta}_{1_i}$ for the four different group compositions in Germany. Here, the values appear to be more or less identical.
\begin{table}[H]
  \caption{The columns provide an overview of the mean values and standard error for $(d_{V})_{i,min}$ and $\hat{\beta}_{1_i}$ for the different group compositions.\label{table:3}}
  \newcolumntype{C}{>{\centering\arraybackslash}X}
    \footnotesize
\begin{tabularx}{\textwidth}{CCCCCC}
  \toprule
     &   \textbf{Female} &  \textbf{Male} &  \textbf{Gender alternating} &  \textbf{Gender random order}  \\
    \midrule
  $\overline{(d_{V})_{i,min}}\pm \sigma$ & $0.31\pm 0.07$     & $0.34\pm 0.08$ &$0.34\pm 0.08$ & $0.36\pm 0.08$ \\
  \\
  $\overline{\hat{\beta}_{1_i}}\pm \sigma$ & $0.96\pm 0.22$    & $0.95\pm 0.23$   &$0.94\pm 0.18$   & $0.90\pm 0.21$   \\
\bottomrule
\end{tabularx}
\end{table}
In addition to obtaining the data shown in \autoref{table:3}, we also compared the different values for $(d_{V})_{i,min}$ and $\hat{\beta}_{1_i}$ using a statistical Tukey test. For $(d_{V})_{i,min}$ and $\hat{\beta}_{1_i}$, 
for each group composition, the p-value is larger than 0.05 so the means are
significantly equal for all four group compositions female, male, gender alternating, and gender random order in Germany. 
\begin{figure}[H]
\centering 
\begin{minipage}[t]{0.49\linewidth} 
\centering
\includegraphics[width=6.5cm]{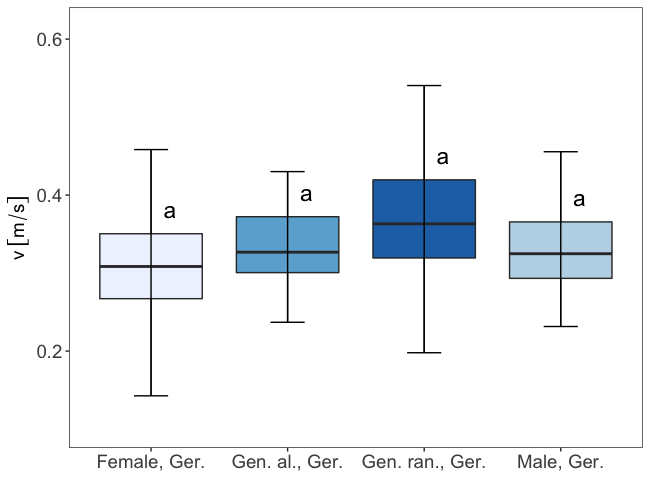}\\
a)
\end{minipage}  
\hfill 
\begin{minipage}[t]{0.49\linewidth} 
\centering
\includegraphics[width=6.5cm]{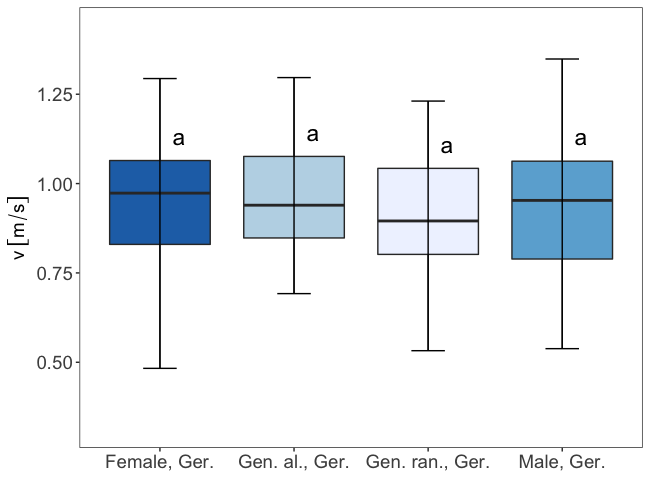}\\
b)
\end{minipage}  
\caption{Boxplots based on a Tukey test for $(d_{V})_{i,min}$ in a) and for $\hat{\beta}_{1_i}$ in b) for different group compositions of female, male, gender alternating, and gender random order in Germany. The same letters above the boxes indicate equality for the mean values between the corresponding groups.}
\label{fig:Verteilung_Dmin_Beta1}
\end{figure}
\autoref{fig:Verteilung_Dmin_Beta1} illustrates the results for $(d_{V})_{i,min}$ in a) and for $\hat{\beta}_{1_i}$ in b) with boxplots based on the Tukey test for different group compositions female, male, gender alternating, and gender random order in Germany. The same letters above the boxes indicate equality of the mean values between the corresponding groups. As the letters above the boxplots are all the same, this indicates that there is equality between the means. As also shown in the previous analysis in this subsection, \ref{Gender differences in distances of neighbouring pedestrians}, it is confirmed that there are no differences between the group compositions at high density. Consequently, the second hypothesis, namely that at high densities, the distance between individuals depends on the gender of the neighboring pedestrians, can not be confirmed.

\subsection{Human factors in fundamental diagrams}
In \cite{paetzke_influence_2022}, the multiple linear regression analysis showed that the headway has the most significant effect on the velocity and other human factors such as gender only have a small effect or can be neglected.\footnote{Further details on the procedure and the structure of the model can also be found in the publication cited.} In this section, we will determine whether taking into consideration additional human factor leads to a more sensitive model. For this purpose, additional factors such as the weight, age, exact height, and gender of the previous pedestrian are taken into account.  

Accordingly, in the new model, the velocity depending on the Voronoi distance, height, gender, age, weight, gender of the previous pedestrian is studied. The variable $gender.prev$ is used for this, and, for all other individual effects, for example, motivation, attention, or excitement which was described in \cite{paetzke_influence_2022} the variable $alloence$ is used. 
It was taken into account that there could be strong correlations between certain human factors. A measurement of the correlation of the factors considered shows obvious dependencies such as the correlation of gender and body height ($p=0.66$), gender and shoulder width ($p=0.71$), or weight and shoulder width ($p=0.75$). In addition, in this analysis, one model is sufficient for all four groups, female, male, gender random order, and gender alternating as the previous results did not show any significant differences between the groups.
Furthermore, the new model was applied to a low density,  $\rho_{gl}\leq 0.75$, and a high density, $\rho_{gl}> 0.75$. Taking into account the results of the previous sections, the low density is the region in which there is no systematic difference between equality and inequality between the mean values of the velocities in the different group compositions.

\begin{figure}[H]
\centering
\includegraphics[width=12cm]{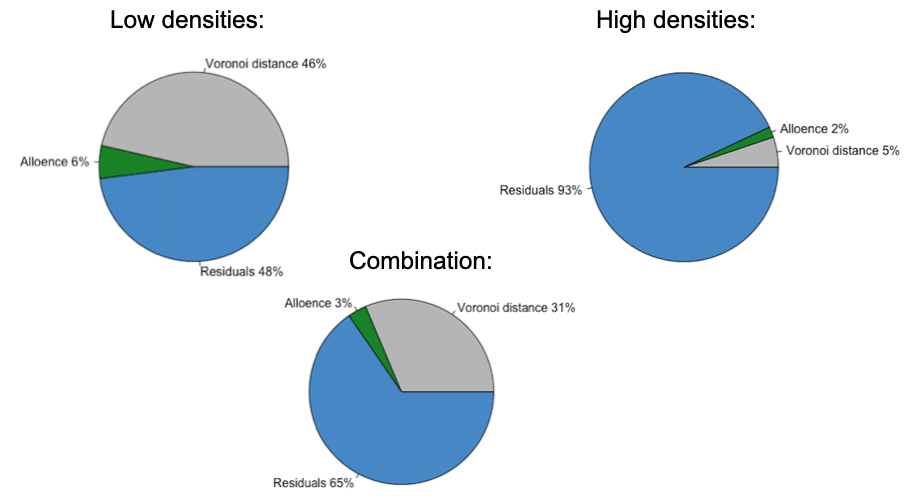}
\caption{Different effects on the individual velocity based on the ANOVA table in pie charts for a low density, $\rho_{gl}\leq 0.75$, a high density, $\rho_{gl}> 0.75$, and for a combination of low and high density.}
\label{fig:Anova_Pie}
\end{figure}

First, the model evaluation using Akaike’s Information Criterion (AIC) is applied to the model introduced in \cite{paetzke_influence_2022}. Step by step, it was decided what factors should be considered in order to obtain the best possible model with the fewest factors without degrading the model. The AIC procedure indicates that gender, age, height, and weight can be omitted. Accordingly, the resulting model is as follows:
\begin{linenomath}
\begin{equation}\label{eq:new_model_end}
 \mbox{Model: $v_{m}=\beta _0+\beta_1 \cdot d_{V_{m}}+\beta_2 \cdot gender.prev_{m}
    +\sum_{i=1}^{N}\beta_{3i} \cdot alloence_{m}+\epsilon_{m}$}\ ,
\end{equation} 
\end{linenomath}

where $m=1,...,n$ and $n$ is the number of all observations of all individuals, $v_m$ is the velocity and $\mbox{alloence}_{m}=1$ for all $m$ belonging to individual $i$ and $0$ for all other $m$. $\beta_{3i}$ is an individual coefficient across all measurement points for each pedestrian.
The new model in \ref{eq:new_model_end} is applied to study the effect of the variables Voronoi distance $d_V$, $\mbox{gender.prev}$, and $\mbox{alloence}$ on the velocity the new model in \ref{eq:new_model_end} is applied. The ANOVA table shows that the p-values for all variables are less than 0.05. This means that all variables considered in the new model have an effect on the individual velocity. \autoref{fig:Anova_Pie} shows the result using pie charts for low and high density and for a combination of these.
Again, the headway has the most effect on the velocity, then all other unknown individual effects. The effect of the gender of the pedestrian in front of the person observed is less than 1$\%$ and can therefore be neglected. Furthermore, the model is better suited to lower densities than to high densities because the velocity is affected more by the Voronoi distance at a low density. At a high density, the effect of the Voronoi distance is decreased, since variations in the range of millimeters occur due to the swaying of the heads. With regard to the third hypothesis, it can be concluded that this hypothesis is false. The analysis provides the same model as before in \cite{paetzke_influence_2022}. When it is taken into account that the gender of the pedestrian in front has an effect, the model is different. However, since the effect is so small, less than 1$\%$, and therefore negligible, the model is identical. 

\section{Conclusions}
The analysis of the single-file experiments conducted by Subaih et al.\,\cite{subaih_experimental_2020} in Palestine showed that in a density range around 1.0~m$^{-1}$ and above, the velocity of groups that are homogeneous with respect to gender is higher than that of heterogeneous groups. In order to test whether this result could be reproduced, further single-file experiments with homogeneous and heterogeneous group compositions were performed in Germany. 

The comparison of different group compositions with respect to gender in different density intervals in the experiment with test persons from Germany could be summarized as follows.
At low densities, the comparison shows no systematic variation, neither with respect to the equality of the mean values of the speed between the group compositions nor with respect to the question of which group is faster. Only in a small density interval around 1.4~m$^{-1}$ do homogeneous groups differ from the heterogeneous groups and show a higher velocity. Between densities of 1.5 to 2.15~m$^{-1}$, the differences lessen and the mean value of the speed of the homogeneous groups approaches the mean of the heterogeneous group gender alternating group. The group gender random order group is the slowest group in this interval. Finally, at high densities, the mean values of the velocities of all groups are equal.

In comparison to the results of the experiments performed in Palestine, there is a certain correspondence but not in every detail. No systematic variation between the homogeneous and heterogeneous groups can be observed for densities lower than 0.8~m$^{-1}$. The difference between homogeneous and heterogeneous groups around a density of 1.4~m$^{-1}$ could be reproduced, but this is less pronounced in the experiments performed in Germany. Moreover, the velocity in the density interval from 1~m$^{-1}$ to 1.5~m$^{-1}$ is higher in Palestine than in Germany. For higher densities, only data for the heterogeneous group in Palestine are available and the mean speed of this group approaches the mean of the German groups. 

Therefore, the first hypothesis, namely that the speed-density relation depends on the gender composition of the group of test persons, is proven to be correct. However, a closer look shows its weak relevance to the experiment in Germany. The difference can only be seen in a narrow density interval and it is small. It cannot be ruled out, however, that the relevance of the effect is stronger in other cultures.

With respect to the relevance of the effect, it should be noted that the verification of the hypothesis depends on the test method as well as on the data preparation. 
Obviously, the size of the binning intervals has an effect on the data. Depending on the size of the individual density intervals selected for different test methods, the systematic variation described above could not be seen. This was already the case for intervals of 0.2. Methods others than the Tukey test, such as the t-test, give similar results. However, tests with a high sensitivity such as the Kolmogorov-Smirnov test, lead to no correspondence of velocity distributions in almost all density intervals and would lead to a rejection of the first hypothesis. For all density intervals, the differences between the means speed of the different group compositions are smaller than the standard deviation. 

For the second hypothesis, a simple linear regression analysis was performed. We used these results to derive the values for the minimal distance $(d_{V})_{i,min}$ and for the reaction time $\hat{\beta}_{1_i}$. A comparison of the mean values of $(d_{V})_{i,min}$ and $\hat{\beta}_{1_i}$ using the Tukey HSD test shows that there are no discernable differences between the four group compositions female, male, gender alternating, and gender random order in Germany. Thus, it can be verified that there are no discernable differences between the reaction time for the different group compositions and the second hypothesis, namely that at high densities, the distance between individuals depends on the gender of the neighboring pedestrians, can not be confirmed.

Finally, we consider the hypothesis on the inclusion of additional human factors that were not previously included such as the weight, exact height, and the gender of the neighboring pedestrian improves the multiple linear regression model developed in \cite{paetzke_influence_2022} is considered. It can be concluded that this hypothesis is false. The analysis provides the same model. Taking into account that the gender of the pedestrian in front has an effect the model makes a difference. However, since the effect is so small, less than 1$\%$, and therefore negligible, the model is identical. 

For further research, the factor of culture could be further investigated. The reason for this is that when comparing the data between Germany and Palestine, we see that the velocity is higher in Palestine. It is unclear where this difference comes from, so more data from other cultures are needed. For Palestine, no data in the higher density range are available to date, so no further conclusions can be derived about the further course of the velocity at present. 
\vspace{6pt} 


\begin{acknowledgements}
This study is based on a one-dimensional single-file experiment that
was performed in 2021 at the Mitsubishi Electric Halle in Düsseldorf,
Germany. Many thanks to Mohcine Chraibi, Anna Sieben, and Mira
Beermann, who all helped with the conceptualization of the study and
with the implementation of the experiments.

The publication costs are funded by the Deutsche Forschungsgemeinschaft (DFG, German Research Foundation) with grant number 49111148.
The experiments were financially supported by the German Federal Ministry of Education and Research (BMBF) within the project CroMa (Crowd Management in Verkehrsinfrastrukturen / Crowd Management in transport infrastructures) under grant number 13N14530 to 13N14533 and by the European Union’s Horizon 2020 research and innovation program within the project CrowdDNA under grant agreement number 899739.
\end{acknowledgements}

\begin{ethics}
  Informed consent was obtained from all subjects involved in the study.

  The data including videos and trajectories used for this study are publicly available in the following archive \cite{DataSingleFileCroMa}.
\end{ethics}

\begin{contributions}
Conceptualization, S.P.; methodology, S.P.; software, S.P.; validation, S.P., M.B. and A.S.; formal analysis, S.P.; investigation, S.P.; data curation, S.P.; writing---original draft preparation, S.P.; writing---review and editing, M.B. and A.S.; visualization, S.P.; supervision, M.B. and A.S.. All authors have read and agreed to the published version of the manuscript.
\end{contributions}

\bibliographystyle{cdbibstyle} 
\bibliography{ped} 

\begin{thebibliography}{10}
\providecommand{\url}[1]{{#1}}
\providecommand{\urlprefix}{URL }
\expandafter\ifx\csname urlstyle\endcsname\relax
  \providecommand{\doi}[1]{DOI~\discretionary{}{}{}#1}\else
  \providecommand{\doi}{DOI~\discretionary{}{}{}\begingroup
  \urlstyle{rm}\Url}\fi

\bibitem{ye_pedestrian_2021}
Ye, R., Zeng, Y., Zeng, G., Huang, Z., Li, X., Fang, Z., Song, W.: Pedestrian
  single-file movement on stairs under different motivations.
\newblock Physica A: Statistical Mechanics and its Applications \textbf{571},
  125849 (2021).
\newblock \doi{10.1016/j.physa.2021.125849}.
\newblock
  \urlprefix\url{https://linkinghub.elsevier.com/retrieve/pii/S0378437121001217}

\bibitem{burghardt_performance_2013}
Burghardt, S., Seyfried, A., Klingsch, W.: Performance of stairs –
  {Fundamental} diagram and topographical measurements.
\newblock Transportation Research Part C: Emerging Technologies \textbf{37},
  268--278 (2013).
\newblock \doi{10.1016/j.trc.2013.05.002}.
\newblock
  \urlprefix\url{https://linkinghub.elsevier.com/retrieve/pii/S0968090X13000946}

\bibitem{cao_pedestrian_2016}
Cao, S., Zhang, J., Salden, D., Ma, J., Shi, C., Zhang, R.: Pedestrian dynamics
  in single-file movement of crowd with different age compositions.
\newblock Physical Review E \textbf{94}(1), 012312 (2016).
\newblock \doi{10.1103/PhysRevE.94.012312}.
\newblock \urlprefix\url{https://link.aps.org/doi/10.1103/PhysRevE.94.012312}

\bibitem{chattaraj_comparison_2009}
Chattaraj, U., Seyfried, A., Chakroborty, P.: {COMPARISON} {OF} {PEDESTRIAN}
  {FUNDAMENTAL} {DIAGRAM} {ACROSS} {CULTURES}.
\newblock Advances in Complex Systems \textbf{12}(03), 393--405 (2009).
\newblock \doi{10.1142/S0219525909002209}.
\newblock
  \urlprefix\url{https://www.worldscientific.com/doi/abs/10.1142/S0219525909002209}

\bibitem{was_effects_2014}
Zhang, J., Tordeux, A., Seyfried, A.: Effects of {Boundary} {Conditions} on
  {Single}-{File} {Pedestrian} {Flow}.
\newblock In: Was, J., Sirakoulis, G.C., Bandini, S. (eds.) Cellular
  {Automata}, vol. 8751, pp. 462--469. Springer International Publishing, Cham
  (2014).
\newblock \doi{10.1007/978-3-319-11520-7\_48}.
\newblock \urlprefix\url{http://link.springer.com/10.1007/978-3-319-11520-7_48}

\bibitem{cao_investigation_2018}
Cao, S., Lian, L., Chen, M., Yao, M., Song, W., Fang, Z.: Investigation of
  difference of fundamental diagrams in pedestrian flow.
\newblock Physica A: Statistical Mechanics and its Applications \textbf{506},
  661--670 (2018).
\newblock \doi{10.1016/j.physa.2018.04.084}.
\newblock
  \urlprefix\url{https://linkinghub.elsevier.com/retrieve/pii/S0378437118305120}

\bibitem{feliciani_empirical_2016}
Feliciani, C., Nishinari, K.: Empirical analysis of the lane formation process
  in bidirectional pedestrian flow.
\newblock Physical Review E \textbf{94}(3), 032304 (2016).
\newblock \doi{10.1103/PhysRevE.94.032304}.
\newblock \urlprefix\url{https://link.aps.org/doi/10.1103/PhysRevE.94.032304}

\bibitem{hu_social_2021}
Hu, Y., Zhang, J., Song, W., Bode, N.W.: Social groups barely change the
  speed-density relationship in unidirectional pedestrian flow, but affect
  operational behaviours.
\newblock Safety Science \textbf{139}, 105259 (2021).
\newblock \doi{10.1016/j.ssci.2021.105259}.
\newblock
  \urlprefix\url{https://linkinghub.elsevier.com/retrieve/pii/S0925753521001041}

\bibitem{ren_fundamental_2019}
Ren, X., Zhang, J., Song, W., Cao, S.: The fundamental diagrams of elderly
  pedestrian flow in straight corridors under different densities.
\newblock Journal of Statistical Mechanics: Theory and Experiment
  \textbf{2019}(2), 023403 (2019).
\newblock \doi{10.1088/1742-5468/aafa7b}.
\newblock
  \urlprefix\url{https://iopscience.iop.org/article/10.1088/1742-5468/aafa7b}

\bibitem{jin_observational_2019}
Jin, C.J., Jiang, R., Wong, S., Xie, S., Li, D., Guo, N., Wang, W.:
  Observational characteristics of pedestrian flows under high-density
  conditions based on controlled experiments.
\newblock Transportation Research Part C: Emerging Technologies \textbf{109},
  137--154 (2019).
\newblock \doi{10.1016/j.trc.2019.10.013}.
\newblock
  \urlprefix\url{https://linkinghub.elsevier.com/retrieve/pii/S0968090X19306527}

\bibitem{cao_fundamental_2017}
Cao, S., Seyfried, A., Zhang, J., Holl, S., Song, W.: Fundamental diagrams for
  multidirectional pedestrian flows.
\newblock Journal of Statistical Mechanics: Theory and Experiment
  \textbf{2017}(3), 033404 (2017).
\newblock \doi{10.1088/1742-5468/aa620d}.
\newblock
  \urlprefix\url{https://iopscience.iop.org/article/10.1088/1742-5468/aa620d}

\bibitem{holl_methoden_2016}
Holl, S.: Methoden für die {Bemessung} der {Leistungsfähigkeit}
  multidirektional genutzter {Fußverkehrsanlagen}.
\newblock Forschungszentrum Jülich GmbH, Zentralbibliothek, Jülich (2016).
\newblock
  \urlprefix\url{https://juser.fz-juelich.de/record/825757/files/IAS_Series_32.pdf}.
\newblock OCLC: 989761703

\bibitem{predtechenskii1978planning}
Predtechenskiĭ, V.M., Milinskii, A.I.: Planning for foot traffic flow in
  buildings.
\newblock Amerind, New Delhi (1978).
\newblock \urlprefix\url{https://books.google.de/books?id=3AZaPwAACAAJ}

\bibitem{weidmann_transporttechnik_1993}
Weidmann, U.: Transporttechnik der {Fussgänger}: {Transporttechnische}
  {Eigenschaften} des {Fussgängerverkehrs}, {Literaturauswertung}.
\newblock Tech. rep., ETH Zurich (1993).
\newblock \doi{10.3929/ETHZ-B-000242008}.
\newblock \urlprefix\url{http://hdl.handle.net/20.500.11850/242008}

\bibitem{vanumu_fundamental_2017}
Vanumu, L.D., Ramachandra~Rao, K., Tiwari, G.: Fundamental diagrams of
  pedestrian flow characteristics: {A} review.
\newblock European Transport Research Review \textbf{9}(4), 49 (2017).
\newblock \doi{10.1007/s12544-017-0264-6}.
\newblock \urlprefix\url{http://link.springer.com/10.1007/s12544-017-0264-6}

\bibitem{fruin_pedestrian_1971}
Fruin, J.J.: Pedestrian planning and design.
\newblock Tech. rep. (1971).
\newblock Fruin1971pedestrian

\bibitem{zhang_pedestrian_2012}
Zhang, J.: Pedestrian fundamental diagrams: comparative analysis of experiments
  in different geometries.
\newblock No.~14 in Schriften des {Forschungszentrums} {Jülich} {IAS}
  {Series}. Forschungszentrum Jülich, Jülich (2012).
\newblock
  \urlprefix\url{https://juser.fz-juelich.de/record/128157/files/FZJ-2012-01052.pdf}

\bibitem{ziemer_mikroskopische_2020}
Ziemer, V.: Mikroskopische {Fundamentaldiagramme} der {Fußgängerdynamik}
  empirische {Untersuchung} von {Experimenten} eindimensionaler {Bewegung}
  sowie quantitative {Beschreibung} von {Stau}-{Charakteristika}.
\newblock Forschungszentrum Jülich GmbH Zentralbibliothek, Verlag, Jülich
  (2020).
\newblock
  \urlprefix\url{https://juser.fz-juelich.de/record/877610/files/IAS_Series_42.pdf}.
\newblock OCLC: 1199751618

\bibitem{ren_contrastive_2019}
Ren, X., Zhang, J., Song, W.: Contrastive study on the single-file pedestrian
  movement of the elderly and other age groups.
\newblock Journal of Statistical Mechanics: Theory and Experiment
  \textbf{2019}(9), 093402 (2019).
\newblock \doi{10.1088/1742-5468/ab39da}.
\newblock
  \urlprefix\url{https://iopscience.iop.org/article/10.1088/1742-5468/ab39da}

\bibitem{subaih_experimental_2020}
Subaih, R., Maree, M., Chraibi, M., Awad, S., Zanoon, T.: Experimental
  {Investigation} on the {Alleged} {Gender}-{Differences} in {Pedestrian}
  {Dynamics}: {A} {Study} {Reveals} {No} {Gender} {Differences} in {Pedestrian}
  {Movement} {Behavior}.
\newblock IEEE Access \textbf{8}, 33748--33757 (2020).
\newblock \doi{10.1109/ACCESS.2020.2973917}

\bibitem{subaih_questioning_2022}
Subaih, R., Maree, M., Tordeux, A., Chraibi, M.: Questioning the {Anisotropy}
  of {Pedestrian} {Dynamics}: {An} {Empirical} {Analysis} with {Artificial}
  {Neural} {Networks}.
\newblock Applied Sciences \textbf{12}(15), 7563 (2022).
\newblock \doi{10.3390/app12157563}.
\newblock \urlprefix\url{https://www.mdpi.com/2076-3417/12/15/7563}

\bibitem{bandini_phase_2010}
Seyfried, A., Portz, A., Schadschneider, A.: Phase {Coexistence} in {Congested}
  {States} of {Pedestrian} {Dynamics}.
\newblock In: Bandini, S., Manzoni, S., Umeo, H., Vizzari, G. (eds.) Cellular
  {Automata}, vol. 6350, pp. 496--505. Springer Berlin Heidelberg, Berlin,
  Heidelberg (2010).
\newblock \doi{10.1007/978-3-642-15979-4\_53}.
\newblock \urlprefix\url{http://link.springer.com/10.1007/978-3-642-15979-4_53}

\bibitem{zhang_universal_2014}
Zhang, J., Mehner, W., Holl, S., Boltes, M., Andresen, E., Schadschneider, A.,
  Seyfried, A.: Universal flow-density relation of single-file bicycle,
  pedestrian and car motion.
\newblock Physics Letters A \textbf{378}(44), 3274--3277 (2014).
\newblock \doi{10.1016/j.physleta.2014.09.039}.
\newblock
  \urlprefix\url{https://linkinghub.elsevier.com/retrieve/pii/S0375960114009517}

\bibitem{migon_favaretto_investigating_2019}
Migon~Favaretto, R., Rosa~dos Santos, R., Raupp~Musse, S., Vilanova, F.,
  Brandelli~Costa, A.: Investigating cultural aspects in the fundamental
  diagram using convolutional neural networks and virtual agent simulation.
\newblock Computer Animation and Virtual Worlds \textbf{30}(3-4) (2019).
\newblock \doi{10.1002/cav.1899}.
\newblock \urlprefix\url{https://onlinelibrary.wiley.com/doi/10.1002/cav.1899}

\bibitem{nguyen_gender-based_2019}
Subaih, R., Maree, M., Chraibi, M., Awad, S., Zanoon, T.: Gender-based
  {Insights} into the {Fundamental} {Diagram} of {Pedestrian} {Dynamics}.
\newblock In: Nguyen, N.T., Chbeir, R., Exposito, E., Aniorté, P., Trawiński,
  B. (eds.) Computational {Collective} {Intelligence}, vol. 11683, pp.
  613--624. Springer International Publishing, Cham (2019).
\newblock \doi{10.1007/978-3-030-28377-3\_51}.
\newblock \urlprefix\url{http://link.springer.com/10.1007/978-3-030-28377-3_51}

\bibitem{dias_pedestrians_2022}
Dias, C., Abdullah, M., Ahmed, D., Subaih, R.: Pedestrians’ {Microscopic}
  {Walking} {Dynamics} in {Single}-{File} {Movement}: {The} {Influence} of
  {Gender}.
\newblock Applied Sciences \textbf{12}(19), 9714 (2022).
\newblock \doi{10.3390/app12199714}.
\newblock \urlprefix\url{https://www.mdpi.com/2076-3417/12/19/9714}

\bibitem{cao_dynamic_2019}
Cao, S., Wang, P., Yao, M., Song, W.: Dynamic analysis of pedestrian movement
  in single-file experiment under limited visibility.
\newblock Communications in Nonlinear Science and Numerical Simulation
  \textbf{69}, 329--342 (2019).
\newblock \doi{10.1016/j.cnsns.2018.10.007}.
\newblock
  \urlprefix\url{https://linkinghub.elsevier.com/retrieve/pii/S1007570418303228}

\bibitem{ma_experimental_2020}
Ma, J., Shi, D., Li, T., Li, X., Xu, T., Lin, P.: Experimental study of
  single-file pedestrian movement with height constraints.
\newblock Journal of Statistical Mechanics: Theory and Experiment
  \textbf{2020}(7), 073409 (2020).
\newblock \doi{10.1088/1742-5468/ab99c0}.
\newblock
  \urlprefix\url{https://iopscience.iop.org/article/10.1088/1742-5468/ab99c0}

\bibitem{zeng_experimental_2019}
Zeng, G., Schadschneider, A., Zhang, J., Wei, S., Song, W., Ba, R.:
  Experimental study on the effect of background music on pedestrian movement
  at high density.
\newblock Physics Letters A \textbf{383}(10), 1011--1018 (2019).
\newblock \doi{10.1016/j.physleta.2018.12.019}.
\newblock
  \urlprefix\url{https://linkinghub.elsevier.com/retrieve/pii/S0375960118312258}

\bibitem{yanagisawa_improvement_2012}
Yanagisawa, D., Tomoeda, A., Nishinari, K.: Improvement of pedestrian flow by
  slow rhythm.
\newblock Physical Review E \textbf{85}(1), 016111 (2012).
\newblock \doi{10.1103/PhysRevE.85.016111}.
\newblock \urlprefix\url{https://link.aps.org/doi/10.1103/PhysRevE.85.016111}

\bibitem{cao_stepping_2018}
Cao, S., Zhang, J., Song, W., Shi, C., Zhang, R.: The stepping behavior
  analysis of pedestrians from different age groups via a single-file
  experiment.
\newblock Journal of Statistical Mechanics: Theory and Experiment
  \textbf{2018}(3), 033402 (2018).
\newblock \doi{10.1088/1742-5468/aab04f}.
\newblock
  \urlprefix\url{https://iopscience.iop.org/article/10.1088/1742-5468/aab04f}

\bibitem{zeng_experimental_2018}
Zeng, G., Cao, S., Liu, C., Song, W.: Experimental and modeling study on
  relation of pedestrian step length and frequency under different headways.
\newblock Physica A: Statistical Mechanics and its Applications \textbf{500},
  237--248 (2018).
\newblock \doi{10.1016/j.physa.2018.02.095}.
\newblock
  \urlprefix\url{https://linkinghub.elsevier.com/retrieve/pii/S037843711830181X}

\bibitem{wang_linking_2018}
Wang, J., Boltes, M., Seyfried, A., Zhang, J., Ziemer, V., Weng, W.: Linking
  pedestrian flow characteristics with stepping locomotion.
\newblock Physica A: Statistical Mechanics and its Applications \textbf{500},
  106--120 (2018).
\newblock \doi{10.1016/j.physa.2018.02.021}.
\newblock
  \urlprefix\url{https://linkinghub.elsevier.com/retrieve/pii/S0378437118300979}

\bibitem{ma_pedestrian_2018}
Ma, Y., Sun, Y.Y., Lee, E.W.M., Yuen, R.K.K.: Pedestrian stepping dynamics in
  single-file movement.
\newblock Physical Review E \textbf{98}(6), 062311 (2018).
\newblock \doi{10.1103/PhysRevE.98.062311}.
\newblock \urlprefix\url{https://link.aps.org/doi/10.1103/PhysRevE.98.062311}

\bibitem{song_experiment_2013}
Song, W., Lv, W., Fang, Z.: Experiment and {Modeling} of {Microscopic}
  {Movement} {Characteristic} of {Pedestrians}.
\newblock Procedia Engineering \textbf{62}, 56--70 (2013).
\newblock \doi{10.1016/j.proeng.2013.08.044}.
\newblock
  \urlprefix\url{https://linkinghub.elsevier.com/retrieve/pii/S1877705813012265}

\bibitem{wang_step_2018}
Wang, J., Weng, W., Boltes, M., Zhang, J., Tordeux, A., Ziemer, V.: Step styles
  of pedestrians at different densities.
\newblock Journal of Statistical Mechanics: Theory and Experiment
  \textbf{2018}(2), 023406 (2018).
\newblock \doi{10.1088/1742-5468/aaac57}.
\newblock
  \urlprefix\url{https://iopscience.iop.org/article/10.1088/1742-5468/aaac57}

\bibitem{fujita_traffic_2019}
Fujita, A., Feliciani, C., Yanagisawa, D., Nishinari, K.: Traffic flow in a
  crowd of pedestrians walking at different speeds.
\newblock Physical Review E \textbf{99}(6), 062307 (2019).
\newblock \doi{10.1103/PhysRevE.99.062307}.
\newblock \urlprefix\url{https://link.aps.org/doi/10.1103/PhysRevE.99.062307}

\bibitem{paetzke_influence_2022}
Paetzke, S., Boltes, M., Seyfried, A.: Influence of individual factors on
  fundamental diagrams of pedestrians.
\newblock Physica A: Statistical Mechanics and its Applications \textbf{595},
  127077 (2022).
\newblock \doi{10.1016/j.physa.2022.127077}.
\newblock
  \urlprefix\url{https://www.sciencedirect.com/science/article/pii/S0378437122001248}

\bibitem{DatenPaperCroMa}
Boomers, A.K., Boltes, M., Adrian, J., Beermann, M., Chraibi, M., Feldmann, S.,
  Fiedrich, F., Frings, N., Graf, A., Kandler, A., Kilic, D., Konya, K.,
  Küpper, M., Lotter, A., Lügering, H., Francesca, M., Paetzke, S.,
  Raytarowski, A.K., Sablik, O., Schrödter, T., Seyfried, A., Sieben, A.,
  Üsten, E.: Pedestrian crowd management experiments: A data guidance paper.
\newblock Collective Dynamics  (2023 (submitted))

\bibitem{ziemer_congestion_2016}
Ziemer, V., Seyfried, A., Schadschneider, A.: Congestion {Dynamics} in
  {Pedestrian} {Single}-{File} {Motion}.
\newblock In: Knoop, V.L., Daamen, W. (eds.) Traffic and {Granular} {Flow} '15,
  pp. 89--96. Springer International Publishing, Cham (2016).
\newblock \doi{10.1007/978-3-319-33482-0\_12}

\bibitem{boltes_automatic_2010}
Boltes, M., Seyfried, A., Steffen, B., Schadschneider, A.: Automatic
  {Extraction} of {Pedestrian} {Trajectories} from {Video} {Recordings}.
\newblock In: Klingsch, W.W.F., Rogsch, C., Schadschneider, A., Schreckenberg,
  M. (eds.) Pedestrian and {Evacuation} {Dynamics} 2008, pp. 43--54. Springer,
  Berlin, Heidelberg (2010).
\newblock \doi{10.1007/978-3-642-04504-2\_3}

\bibitem{BOLTES2013127}
Boltes, M., Seyfried, A.: Collecting pedestrian trajectories.
\newblock Neurocomputing \textbf{100}, 127--133 (2013).
\newblock \doi{https://doi.org/10.1016/j.neucom.2012.01.036}.
\newblock Special issue: Behaviours in video

\bibitem{liao_detection_2016}
Liao, W., Tordeux, A., Seyfried, A., Chraibi, M., Zheng, X., Zhao, Y.:
  Detection of {Steady} {State} in {Pedestrian} {Experiments}.
\newblock In: Knoop, V.L., Daamen, W. (eds.) Traffic and {Granular} {Flow} '15,
  pp. 73--79. Springer International Publishing, Cham (2016).
\newblock \doi{10.1007/978-3-319-33482-0\_10}

\bibitem{dataSubaih}
{Forschungszentrum Jülich}: Influence of gender in single-file movement.
\newblock \url{http://ped.fz-juelich.de/da/2018singleFile}.
\newblock \doi{10.34735/ped.2018.5}

\bibitem{DataSingleFileCroMa}
{Forschungszentrum Jülich}: Oval {{Experiment}}: {{Single File Motion}}.
\newblock Pedestrian Dynamics Data Archive  (2022).
\newblock \doi{10.34735/PED.2021.5}.
\newblock \urlprefix\url{http://ped.fz-juelich.de/da/2021oval}

\end{thebibliography}

\end{document}